# Techniques, advances, problems and issues in numerical modelling of landslide hazard.

**Th.W.J. van Asch[1], J.-P. Malet[1], L.P.H. van Beek[1], D. Amitrano[2]**

1. Faculty of Geosciences, UCEL, Utrecht University, Heidelberglaan 2, Po.Box 80.115, 3508 TC Utrecht, The Netherlands.
2. LIRIGM-LGIT, Université Joseph Fourrier, Maison des Géosciences, 1381 rue de la Piscine, BP53, F-38041 Grenoble Cedex 9, France.

*. Corresponding author: t.vanasch@geo.uu.nl

**Abstract:**
Slope movements (*e.g.* landslides) are dynamic systems that are complex in time and space and closely linked to both inherited and current preparatory and triggering controls. It is not yet possible to assess in all cases conditions for failure, reactivation and rapid surges and successfully simulate their transient and multi-dimensional behaviour and development, although considerable progress has been made in isolating many of the key variables and elementary mechanisms and to include them in physically-based models for landslide hazard assessments. Therefore, the objective of this paper is to review the state-of-the-art in the understanding of landslide processes and to identify some pressing challenges for the development of our modelling capabilities in the forthcoming years for hazard assessment. This paper focuses on the special nature of slope movements and the difficulties related to simulating their complex time-dependent behaviour in mathematical, physically-based models. It analyses successively the research frontiers in the recognition of first-time failures (pre-failure and failure stages), reactivation and the catastrophic transition to rapid gravitational processes (post-failure stage). Subsequently, the paper discusses avenues to transfer local knowledge on landslide activity to landslide hazard forecasts on regional scales and ends with an outline how geomorphological investigations and supporting monitoring techniques could be applied to improve the theoretical concepts and the modelling performance of physically-based landslide models at different spatial and temporal scales.

## 1. Introduction

The sustainable development of mountainous areas and safety to the citizens require sophisticated and reliable analyses of hazardous processes and consequent risks. A major threat is induced by all types of slope movements (e.g. falls, topples, slides, lateral spreads, flows) which are triggered in these areas and which represent one of the most destructive natural hazards on earth (Brabb, 1991). Human casualties are important and economical losses may reach 1 or 2% of the gross national product in many developing countries (Schuster & Highland, 2001). As stated in 2006 by the United Nation University, Asia suffered 220 landslides in the past century – by far the most of any world region – but those in North, Central and South America have caused the most deaths and injuries (25,000+) while Europe's are the most expensive – causing average damage of almost $23 million per landslide – (OFDA/CRED, 2006). Moreover, slope movement activity seems to increase because of global warming and anthropic actions (Schuster, 1996). Analysing, evaluating and mitigating the hazard and risk associated to slope movements is therefore a challenge for many earth scientists, engineers and decision-makers.

Forecasting of both the spatial and temporal probability of occurrence and the intensity of all types of slope movements is a necessary task to characterize and analyse quantitatively the hazard in an area (Aleotti & Chowdhury, 1999; Bonnard *et al.*, 2003). Occurrence probability is the probability that a certain phenomenon will occur over a specified period of time. It can be evaluated in both qualitative and quantitative terms. The spatial probability of occurrence (that is by definition a non-temporal concept) is called susceptibility. Intensity is a measure of the destructive potential of a phenomenon, based on a set of physical parameters, such as velocity, thickness of the displaced debris, volume, energy and impact forces. Again, intensity can be expressed in both qualitative and quantitative terms. Intensity varies with location along and across the travel path of the material and therefore it should ideally be described using a spatial distribution.

At the present state of knowledge, understanding, forecasting and controlling the hazard associated to slope movements is still an empirical task. It requires mixing qualitative and quantitative analyses of data (including model simulations) proceeding from distinct disciplines (geomorphology, geology, hydrology, hydrogeology, geophysics, geotechnics, civil engineering). Analysis can be performed at several spatial and temporal scales according to the objective of the hazard assessment (Aleotti & Chowdhury, 1999; van Westen *et al.*, 2005). Accordingly the methods and tools used for the analysis are radically different, ranging from empirical or statistical techniques (e.g. multivariate analysis) which are generally applied to predict slope susceptibility at regional scale to more process-based approaches (e.g. limit-equilibrium methods, numerical deformation methods) which are applied at the local scale.

The purpose of this article is to present the techniques, advances, problems and likely future developments in numerical modelling of landslide hazard at the local scale with a focus on physically-based (or process-based) approaches. The paper does not review the various techniques and methods available at the regional scale for susceptibility and hazard evaluation (see for instance Guzzetti *et al.*, 1999; van Westen, 2001).Moreover, the focus is restricted on the analysis of the literature concerning slide and flow processes.

• First, the special nature of slope movements is discussed and hence some of the difficulties related with simulating their complex time-dependent behaviour in numerical models. The utility of numerical modelling in understanding slope movement mechanism and in simulating scenarios for hazard assessment is also outlined;

• Second, some of the major problems in analysing the mechanisms at the local scale are summarized on the basis of an extensive literature review, and some research directions are proposed. The authors distinguish the models according to the temporal stage of development of the processes (pre-failure stage, failure stage, post-failure stage);

• Third, up scaling of the site-scale knowledge to the forecasting of the hazard at the catchment or regional scale is discussed;

• Finally, ways of improving (as well as quantitatively evaluating) our modelling performance are proposed.

## 2. Special nature of slope movements and available numerical methods

### *A variety of slope movements and controlling factors*

In both the geomorphological and the geotechnical literature (Dikau *et al.*, 1996; Cruden & Varnes, 1996), the term slope movement characterizes all varieties of ground failure and downslope movement of earth material controlled by gravity. One of the simplest methods of classification (Fig. 1) is that initially proposed by Varnes (1978) and extended by Vaunat *et al.* (1994) and Leroueil (2001): they classify slope movements according to the type of movement, the type of material and the movement phase or state of activity (e.g. the rate of development over a period of time, Dikau *et al.* 1996).

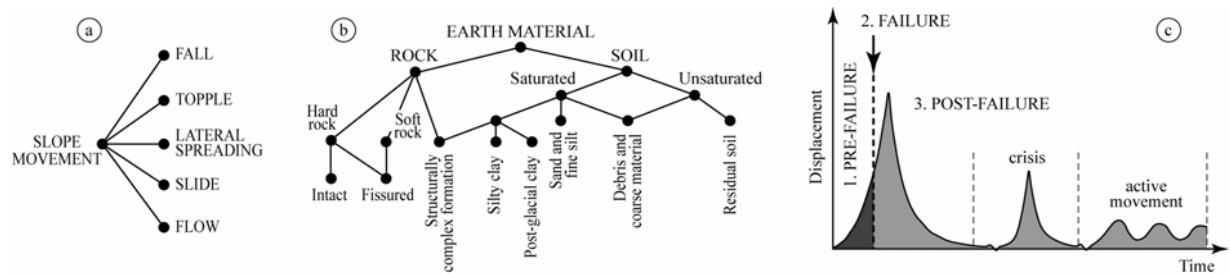

Fig. 1. – Classification of slope movements according to the movement types (1a), the material types (1b) and the movement phases (1c) (modified from Vaunat *et al.*, 1994).

Types of movements are distinguished according to the geomorphological classification proposed by Cruden and Varnes (1996) and Dikau *et al.* (1996) which assume five principal types. The definition of each type of movement is given in Table 1.

Tab. 1 – The five principal types of slope movement

| Type of movement | Definition |
|---|---|
| Fall | A slope movement for which the mass in motion travels most of the distance through the air, and includes free fall movement by leaps and bounds and rolling of fragments of material. A fall starts with the detachment of material from a steep slope along a surface on which little or no shear displacement takes place. |
| Topple | A slope movement that occurs due to forces that cause an over-turning moment about a pivot point below the centre of gravity of the slope. A topple is very similar to a fall in many aspects, but do not involve a complete separation at the base of the failure. |
| Lateral spreading | A slope movement characterized by the lateral extension of a cohesive mass over a deforming mass of softer underlying material in which the controlling basal shear surface is often not well-defined. |
| Slide | A slope movement by which the material is displaced more or less coherently along a recognisable or less well-defined shear surface or zone. |
| Flow | A slope movement characterised by internal differential movements that are distributed throughout the mass and in which the individual particles travel separately within the mass. |

These five principal types may sometimes be combined or may be cascaded in time and space, resulting in 'complex movements', which consists of more than one type (e.g. a rotational-translational slide) or those where one type of movement develops into a second type (e.g. slump-earthflow). (DA: this is the case of Supersauze isn't) Many slope movements are complex, although one type of movement generally dominates over the others at certain areas within the displaced mass or at a particular time. Thus, except for falls and topples, many slope movements are not first-time movements; it is therefore important to identify the initial phenomena in order to study properly the present movement patterns.

Another important qualitative criterion in identifying the type of slope movements is to characterize their size and especially their thickness. Varnes (1984) categorises the slope movements as shallow (if the thickness is less than 2m thick), medium (if the thickness is comprised between 2 and 5m) and deep-seated (if the thickness is larger than 5m). The thickness is difficult to estimate from the surface using solely geomorphological criteria and it calls for specific investigation.

The behaviour of slope movement is highly controlled by the nature of the geological material and so the physical, hydrological and geotechnical properties have to be established. In the geotechnical classification (Fig. 1b), the types of material are gathered into ten main classes: hard intact rocks, hard fissured rocks, soft rock, structurally complex formations, stiff clays, post-glacial clays, silts and fine sands, debris and coarse materials, truly collapsible soils (loess, etc) and other unsaturated materials such as residual soils. The material is largely discontinuous, anisotropic, inhomogeneous and behave as a non-elastic medium because it is continuously under stress and loaded by dynamic movements (DA comments: this is not the cause of the non elastic behaviour. If I understood what you mean, the viscous part of the material behaviour is sollicitated because of the temporal changes in stress an loading. I suggest to change the last sentence) (tectonic movement, earthquake, land uplifting/subsidence, glaciation cycle, tide). The material is also a fractured porous medium containing fluids (in either liquid or gas phases), under complex in situ conditions of stresses, temperature and fluid pressures. The complex combination of constituents and its long history of formation make therefore the material behaviour difficult to represent via a mathematical model.

Finally, the behaviour of slope movement is time-dependent (Flageollet, 1996; Qin *et al.*, 2001), and the movement phase is split into pre-failure, failure and post-failure stages (Tab. 2) with the possibilities of occasional reactivation. All types of movements at a given stage are associated with specific control variables that are divided in predisposition factors and triggering factors (DA adds: note that the limit between these two kinds of factors is not sharp and is sometime controversous).

Tab. 2 – The three phases of movement and their associated specific control variables.

| Movement stage | Definition |
| --- | --- |
| Pre-failure | This stage of movement includes all the deformation processes leading to failure. The processes are controlled by predisposition factors that give information about the present situation and determine the initial or boundary conditions influencing the values of local (triggering) variables. |
| Failure | This stage takes place at the moment of the full development of a continuous (localized or diffuse) shear surface through the entire mass and is generally reflected in an increase of the displacement rate. Failure can occur along discrete pre-exiting surfaces or because of global deformation of a massif. This stage is controlled by the occurrence of a triggering factor. |
| Post-failure | This stage takes place at the moment of the mass detaches from the massif with an initial acceleration. The post-failure stage ends when the mass of stops definitively or temporarily. The post-failure stage can reflect several situations: the movement can be slow and of long-duration, or very rapid and of quick-duration, and may involve occasional reactivation and crisis (e.g. acceleration of the rate of movement due to a significant modification of the predisposing and triggering factors). |

Movement is resisted by the shear strength of the material (cohesion and total inter-particle friction) that can be mobilised along the slip surface. Hence, the ratio between the maximum of available shearing resistance (resisting forces) and the shear stress (disturbing forces), or safety factor F, is a measure for the stability of a slope, and when the two cancel each other, F=1, failure is imminent (DA adds: or is achieved ) (Bromhead, 1992; van Beek, 2002). From the physical point of view, it may be useful to visualize slopes as existing in one of the following three stages (Fig. 2):

• Stable, F>1.5: the margin of stability of the slope is sufficiently high to withstand all destabilising forces;

• Marginally stable, 1.0<F<1.5: the slope is likely to fail at some time in response to destabilising forces reaching a certain level of activity;

• Actively unstable, F<1.0: destabilising forces produce continuous or intermittent movements.

DA commments: theroretically, F can not be lesser than one (equal to one in worse cases) as it is the ratio between the material strength and the actual loading of the material (which can not be higher than its strength), in the same manner a yield criterion of a given material can not be surpassed but only reached. I suggest to correct this part of the text (juste write F~=1) and the next figure or to precise that this case can only be found during short periods (earthquakes or high rain infill). When F reaches 1, the adjustment with the loading is realized by the viscous component part of the behaviour if any, and result in an acceleration of the strain.

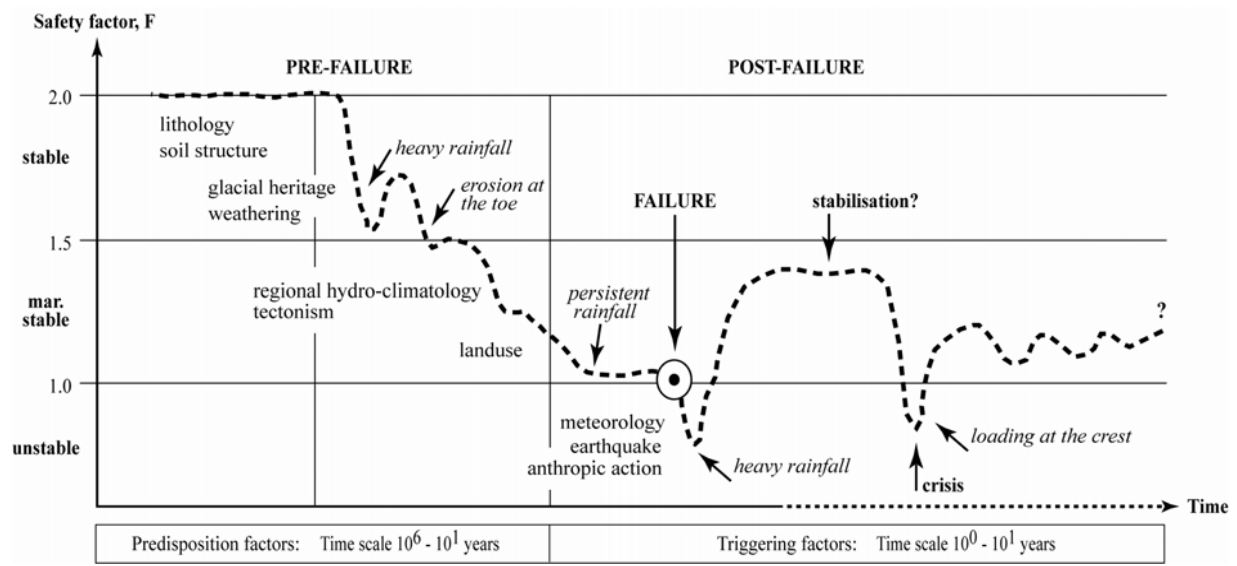

Fig. 2 – Time influence of predisposition and triggering factors on instability.

These three stages provide a useful framework for understanding the causal factors of slope movements. It can be seen that short-term variations in safety factor may occur due to seasonal variations in groundwater levels and pore pressures, while longer term trends may reflect the influences of weathering or glaciation cycles or longer term changes in regional groundwater conditions. This approach is useful in emphasising that slope movements may not be attributable to a single control factor, but always to a combination of predisposition and triggering factors. Thus, the predisposition factors change most times only gradually over time whereas the triggering factors are transient (Crozier, 1986; Leroueil, 2001). Even if it is often difficult to distinguish the true cause of a slope movement, there are a number of triggers which may be operating either to increase the shear stress or to decrease the shearing resistance of the material (DA adds: eg: by decreasing the normal shearing because of pore pressure increase). There are also causes affecting simultaneously both terms of the safety factor. Table 3 presents the most common triggering factors of slope movements.

Tab. 3 – Common triggering factors of slope movements.

| |
|---|
| *Increase in shear stress* |
| • Erosion and excavation at the toe of the slope |
| • Subterranean erosion (piping) |
| • Surcharging and loading at the crest (by deposition or sedimentation) |
| • Rapid drawdown (man-made reservoir, flood, high tide, breaching of natural dams) |
| • Earthquake |
| • Volcanic eruption |
| • Modification of slope geometry |
| • Fall of material (rock and debris) |
| *Decrease in shearing resistance* |
| • Water infiltration (rainfall, snowmelt, irrigation, leakage of drainage systems) |
| • Weathering (freeze and thaw weathering, shrink and swell weathering of expansive soils) |
| • Physico-chemical changes |
| • Fatigue due to cyclic//static loading and creep |
| • Vegetation removal (by erosion, forest fire, drought or deforestation) |
| • Thawing of frozen soils |
| *Possible increase in shear stress and decrease in shearing resistance* |
| • Earthquake shaking |
| • Artificial vibration (including traffic, pile driving, heavy machinery) |
| • Mining and quarrying (open pits, underground galleries) |
| • Swinging of trees |

***Strategy for hazard assessment: concepts, available modelling tools and numerical methods***

Although some qualitative answers to important questions can be made using best engineering or best geomorphologic judgment, in many instances human reasoning alone is inadequate to synthesize the conglomeration of factors involved in analyzing complex slope stability problems *(*van Westen *et al.*, 2005). The best tool to help slope stability engineers meet the challenge of analyzing and forecasting the hazard is a mathematical model describing the relations between the predisposing and triggering causes (model inputs) and the responses of the slopes (model outputs), either explicitly or implicitly. However in many cases, building an effective computer model is very difficult due to the specific nature and the 4-D pattern (DA adds: space and time) of slope movements, as outlined in the former section.

Hazard assessment supposes the identification of a slope concept and of the elementary mechanisms controlling instability. Mechanisms of slope movements are fundamental for a regionalisation of the knowledge and hence a quantitative hazard assessment at several scales (Brunsden, 1999). Mechanisms of slope movements are idealized ways that slope material might move (Hutchinson, 1988). Hence, each different mechanism requires a different type of hydro-mechanical analysis, but there may be several hydro-mechanical analyses addressing the same mechanism. A synthesis of the fundamental failure mechanisms (e.g. sliding over an existing surface or across a shear zone at the base of the moving mass, progressive plastic deformation of the material and formation of a shear zone, rotation of the principal stresses at the base of the moving mass, static liquefaction, undrained loading: Hutchinson & Bhandari, 1971; Sladen *et al.*, 1985; Urcioli, 2002) and fundamental post-failure mechanisms (e.g. grain-crushing, rheofluidification, vibration energy; Sassa 1998; Iverson *et al.*, 1997), described and analysed theoretically or numerically, can be found in Terzaghi (1950), Morgenstern & Sangrey (1978), Hutchinson (1988, 1993) and Picarelli *et al.* (2000).

To adequately represent the slope movements in computer models, it is necessary to include the following features in the model concept:

• the local characterization of the geometry and internal structure (e.g. layering, discontinuities) of the slope (the so-called geometrical model) in order to define the probable extension of the slope movement;

• the local identification of eye-witness evidences of movements (e.g. kinematic fractures, lobes, scarps, horst/graben structures, etc) in order to define the probable state of activity of the slope movement (the so-called geomorphological or morphostructural model). This model is constructed through photo-interpretation of multi-source documents and field work;

• the monitoring of key controlling variables (e.g. rainfall, air and soil temperature, soil moisture content, water level at many points, surfacial and in-depth displacement at many points, etc) at a time frequency in accordance with the rate of movement in order to define the kinematics of the slope movement (the so-called kinematic model). This point consists also in characterizing the pre-existing state of stress within the slope;

• the characterization of the physical and hydro-mechanical properties (e.g. constitutive laws) of the geologic material at different locations (the material is inhomogeneous), in different directions (the material is anisotropic), at different scales (the material is scale-dependent) and for many stress configurations (the material in non-elastic and may undergo creep or plastic deformation DA adds: depending on the stress amplitude) in order to define time/rate-dependent behaviour of slope movement (the so-called geotechnical model);

• the cross-analysis of all these data and the merging of the sub-models in order to identify the probable mechanism of movement (the so-called geomechanical model) with the associated relevant physical processes (and their mathematical representations, key variables and parameters).

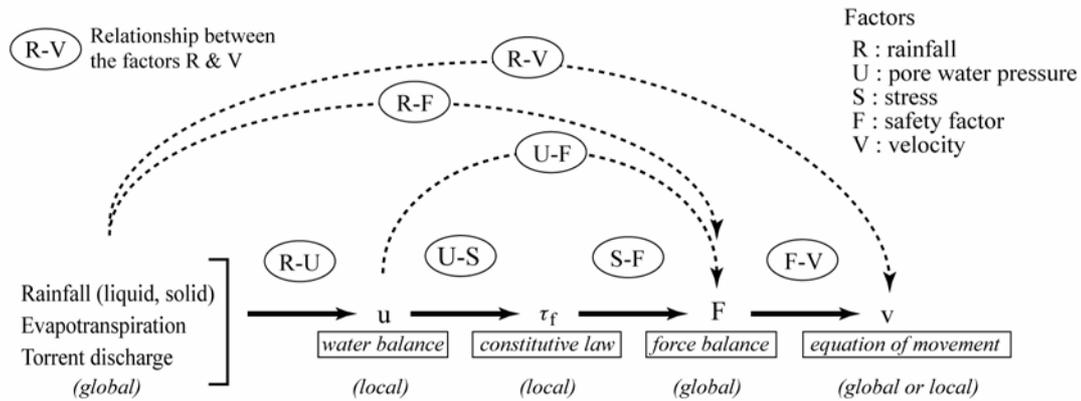

Fig. 3 – Relations simulated in a physically-based landslide model (in Malet, 2003; modified from Leroueil, 2002).

The figure 3 explains the relationships between the controlling variables and thus the constitutive mathematical equations of a process-based landslide model. The figure highlights the relations for rainfall-induced landslides which dynamic is simulated by a hydrological model (R-U relationship) coupled to a mechanical model (U-S-F-V relationship) by a material constitutive law embedded in an equation of motion (DA adds: if the explicit approach is used).

The extent to which these features are captured for the investigated slope as well as the scope of the work, will (partly) determine the numerical modelling approach to be used, the space and time conditions of the analysis (2D or 3D, constant or time-dependent simulations), and the reference to a discontinuous or equivalent continuum approach. The main problem with the discontinuous approach is to determine the location and geometry of the natural discontinuities, while the main problems with the continuum equivalent approach is the evaluation of the hydro-mechanical properties of the geologic material which cannot be determined in the laboratory.

To discretize the time and space dimensions, the most widely used numerical methods are the limit-equilibrium method (LEM), the finite element method (FEM), the boundary element method (BEM) and the finite difference method (FDM) for the continuum equivalent approach, and the discrete element method (DEM) for the discontinuous approach. LEMs do not allow the evaluation of stress and strain conditions in the slope, so they are not able to reproduce the crucial role played by deformability slope movements (Bromhead, 1996; Griffiths & Lane, 1999). FEMs and FDMs, on the one hand, are the most flexible methods because of their ability in handling material heterogeneity, non-linearity and boundary conditions, but due to their interior discretization, they cannot simulate infinitely large domains (DA comments: this is the case for all the methods) and the computation time can be very important. BEMs, on the other hand, require discretization at the boundary of the solution domains only, thus greatly simplifying the input requirements, but are not convenient when more than one material must be taken into account DA adds: neither for integrating material spatial heterogeneity. It is the most efficient technique for fracture propagation analysis. DEMs represent a discontinuous medium as assemblages of blocks formed by connected fractures in the problem domain, and solve the equations of motion of these blocks through continuous detection and treatment of contacts between the blocks. Handling large displacements including fracture opening and complete detachments is therefore straightforward in these methods.

Hence, both the modelling and any subsequent scenario simulations will always contain subjective judgements. On the author's points of view, the challenge is now not on developing truly fully (thermo-hydro-mechanical) coupled numerical models which require detailed knowledge of the geometrical and physical properties and parameters of the slopes, but to clarify how to develop an adequate numerical model DA adds: for answering a precise question (how when where, what size etc.). The model does not have to be complete and perfect: it only has to be adequate for the purpose of hazard assessment.

*Scope of the paper*

As stated here above, the purpose of this paper is to make a review of the state-of-the-art in our understanding of slope movements, and to indicate some challenges for the development of our modelling capabilities in the forthcoming years. The paper uses recent available reviews of slope stability models (Bromhead, 1996; Brunsden, 1999; Vulliet, 2001; Vulliet & Dewarrat, 2001) and recent Conference Proceedings on landslides and debris flows since 10 years (*7th ISL*, Trondheim, 1996; *8th ISL*, Cardiff, 2000; *9th ISL*, Rio de Janeiro, 2004; *2nd DFHM* Conference, Taipei, 2001; *3rd DFHM* Conference, Davos, 2003) as well as recent research papers published in International Journals. Writing a state-of-the-art paper in a few pages has conducted to some choice, often subjective, and there exist of course some lacks in our review. Only papers describing conceptual, theoretical or numerical studies are referred in our review; papers dealing with detailed case studies are not within the scope of this paper.

The paper analyses successively the major research frontiers in the understanding of first-time failures (pre-failure and failure stages) and in the understanding of post-failure behaviours highlighting critical factors in the propagation of slow-moving and rapid gravitational processes. Then procedures for landslide hazard assessment and mapping at the catchment or region scales are discussed, and ways of improving our modelling performance (geomorphology, monitoring techniques) are proposed.

### 3. Pre-failure preparatory stage (or behaviour) – To be completed by DA

State-of-the-art in understanding the preparatory mechanisms (weathering, tectonic and fracture development, etc).

One of the main issue for modelling is to forecast first time failure of landslides. In order to do this properly we need to investigate the long-term preparatory evolution of the slope and the final triggering mechanisms. The long term evolution is related to the rate of chemical and mechanical weathering of the rock and soil material weakening of the strength and the chemical (changing) condition of the groundwater near potential slip-surface. Also of great importance is the geomorphological evolution of the slope which is determined by a variety of processes working with different intensity depending on the climatic conditions. Erosion and sedimentation processes on the slope and external erosion at the foot are loading and unloading processes which change the equilibrium system of the slope *(Brunsden 1998)*. To forecast first time failure in rock material it is important to follow the evolution of the fractural system which is inherited by the tectonic history in rocks. Thermo mechanical and chemical processes induced by circulating water in the preferential fissure system will further develop the fracture system. *(Peng 1973, Boukharov et al. 1995, Kilburn and Petley 2003)* Most efficient is the freeze thaw cycles of water in the fissure system. The circulating water will provoke chemical changes and also the varying temperature and pore pressures will affect the system of fissures and discontinuities leading to changes in the stress strain field and possible failure.

Apart from this external influences one has to consider the growth of cracks in rocks due to load and unloading mechanisms which controls the path to failure. The development of individual cracks at the micro scale can be considered as well as population of cracks at the meso scale *(, Kranz 1980, Masuda 2001, Amitrano et al 1999, Amitrano 2003, Amitrano, 2006).*

The damage localization process in rocks has been often modeled considering either a discontinuous media containing propagating cracks (Costin, 1983; Cowie et al., 1993; Scavia, 1995) or a continuous material subject to a bifurcation phenomena (Rice, 1975) with some applications to slope instability (e.g., Scavia and Castelli, 1996). An intermediary approach consists in considering the material to be continuous at mesoscale. The cracking is taken into account through elastic damage (reduction in the apparent elastic modulus). In this way, it is possible to model either macroscopic plasticity (Zapperi et al., 1997) or macroscopic brittleness (Tang, 1997; Tang and Kaiser, 1998). Some applications of this approach have been developed mainly for underground mining failure. Amitrano et al. (1999) proposed a model able to switch continuously from macroscopic plasticity, with diffuse damage, to macroscopic brittleness, with localised damage. These numerical results appear to be in good agreement with laboratory experiments and earthcrust observations (Amitrano, 2003), but have not been applied to rocky slope instabilities until recently. Following this mesoscale approach and considering the subcritical growth of a cracks population, Amitrano and Helmstetter (2006) simulated the brittle creep phenomenom, i.e. the three stages of creep (primary, secondary and tertiary creep) associated to different stages of damage spatial distribution (diffuse to localized). The tertiary creep appears to be associated to both strain and seismicity acceleration in accordance with in-situ observations of muddy slope failures (Voight, 1988; Petley et al, 2002) and of forerunners of a chalk cliff collapse (Amitrano et al, 2005).

The figure xx present an application of this modelling to an idealized rocky slope (from Amitrano, 2005) considering an initially intact material.

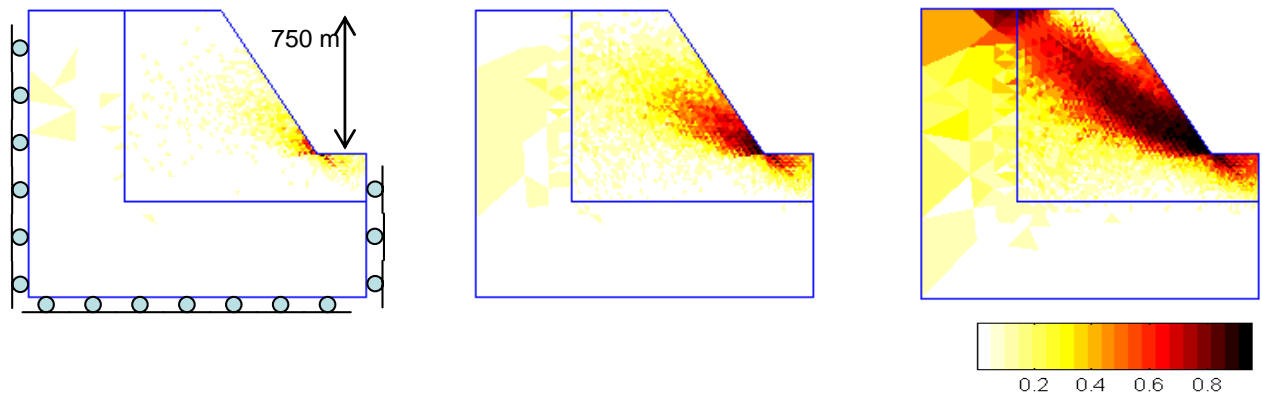

Figure xx: Modelling of the progressive damage of an idealized rocky slope based on brittle creep theory (from *Amitrano, 2005*).

This simulation shows the propagation of damage within the slope and the final state of damage, which could be compared with images provided by, e.g., geophysical prospecting. For reaching a better representation of real slope this modelling should integrate tectonics setting, erosion history and real material heterogeneity.

From theoretical and experimental point of view, the life time and strain rate during brittle creep is known to be sensible to the temperature, water saturation and effective pressure (Scholz, 1968, 1970; Kranz, 1980; Kranz et al. 1982; Masuda, 2001). As the water flow is related to the state of damage through the permeability, damage propagation and localization and water flows within the damaged material can be related by a positive feedback what can be a supplementary explanation of the water dependence of the strain rate observed for landslides including rocky slopes.

The damage process can also be characterized by a hydro-chemical signature as the increase of free surface induces a higher reactivity of the rock/water interface. This theoretical consideration is in agreement with laboratory observation (Ojala et al., 2003; Bruderer-Weng et al., 2004; Song et al., 2004) and provides a possible mechanical interpretation of in-situ observation of the correlation between slope deformation and chemical composition of flowing water (Binet et al. 2004, Charmoille et al., 2005).

## Examples of models, challenges & future research directions

On a shorter time scale more precise prediction of failure can be made in the last stage by monitoring the displacement of the slope. *(Bhandari 1988; Zvelebil and Moser 2001; Petley et al., 2002; Petley et al., 2005).* Important for forecasting the time of failure is the acceleration phase in the displacement which can be described by a power law *(Saito and Uezawa, 1961; Kennedy and Niermeyer 1971; Voight 1989)* or an exponential law *(Petley et al. 2002).* The same kind of acceleration has been observed for the microseismicity induced by crack propagation before the collapse of a chalk cliff (Amitrano et al, 2005). Various authors try to explain the character of the curves by creep processes in the material which were measured in lab experiments, by concepts of damage model in fractured rock (*Voight 1989, Amitrano and Helsmstetter, 2006*) and by a slider block model (*Scholz, 1998, Helmstetter et al. 2004*). The problem as always is to transfer these concepts and models which were developed in the laboratory to the real world. There we are confronted with external perturbations in the evolution to failure (climatic, hydrologic, tectonic, seismic waves, human-induced (*Fukuzono 1985; Voight 1989; Qin et al. 2001),* which makes it more difficult to quantify the failure path. Different techniques (NDS: nonlinear dynamical systems techniques; ANNs: artificial neural networks) have to be further developed in the field to analyse and to forecast the influence of these external factors (rain, fluid pore pressure) on the failure pattern of these slopes *(Mayorraz et al. 1996; Mayorraz and Vulliet, 2002*). This will improve our ability to identify the pattern of failure and to understand the physical mechanisms behind the movement of a potentially unstable hill slope.

Following the previous review, it appears that a mutliphysics modelling associating damage, water flow and chemical exchanges could be realized in order to provide physical interpretation of the in-situ observations, particularly for the search of possible forerunner behaviours useful for the forecast of slope failure. It is therefore important to develop laboratory and controlled field experiments coupled to the development of numerical models which describes the changing stress distribution in rocks imposed by external loads, changing fluid pressure and temperature changes chemical erosion and especially freeze thaw cycles in relation to the development of discontinuities.

## 4. Failure behaviour: the role of hydrology as a dynamic trigger

### *The necessity of understanding the hydrological triggering mechanisms*

Worldwide rainfall-triggered landslides occur more frequently than earthquake-triggered landslides. In general terms, infiltration and the resultant transient changes in the hydrological systems is the most common trigger of landslides (van Asch *et al.*, 1999). There are many main types of hydrological triggering mechanisms (dependent on the state of the system) which control the threshold for first-time failure, but also for landslide reactivation and the progress of movement.

The more well known triggering system occurring in shallow as well as deeper landslides is related to an increase in pore pressures resulting in a decrease in effective stress and strength, and thus a possible decrease the factor of safety to 1.0. However, water infiltration may have other effects both before and after slope failure. Especially, on steep slopes in shallow soils, landslips can be triggered by infiltrating water reducing the increase in effective stress due to matric suction but without generating positive pore pressures (van Asch & Sukmantalya, 1993; Terlien, 1995; Sun *et al.*, 1998; Brooks *et al.*, 2004). Figure 4 outlines this mechanism for a slope profile in Indonesia? (Theo, explain Figure). This mechanism may be of paramount importance in cohesionless soils (Fredlund *et al.*, 1996). Another important but quite different hydrological trigger is the water table rising in infinite slopes causing a rotation of the principal stresses at the base of the soil, and consequently of the potential failure planes (Picarelli *et al.*, 2000; Urcioli, 2002). Finally, the water run-off system in steep catchments experiencing high discharges can trigger debris flows (Blijenberg, 1998) or conduct to an extra-infiltration rate in a landslide body.

Fig. 4 – A conceptual model of transient percolating water in a unsaturated soil profile of what type of soil, where ??. The decrease in matrix suction over time (and increase in soil moisture with depth) does not necessary imply an equivalent decrease in effective stress, but may conduct to failure in steep or cohesionless slopes.

Surprisingly, modelling these hydrological triggering factors has been curiously slow to gain acceptance (Bromhead & Dixon, 1984; Anderson and Kemp, 1988), compared to rainfall-runoff modelling or catchment hydrological modelling. Nevertheless, this aspect has been gaining ground in recent years as geotechnical, geomorphological and hydrological models are drawn closer together (Picarelli *et al.*, 2005). The research frontiers are connected with the complexity of real landslides, the difficulty to monitor groundwater levels or soil moisture contents in 'moving environments', the difficulty to understand the water pathways within the landslide bodies (Brunsden, 1999). Many authors (Okunushi & Okumura, 1987; Haneberg, 1991; van Asch *et al.*, 1999) have shown that the quality of the hydrological model had a greater influence on the general behaviour model than the geomechanical model.

Consequently, the occurrence of rainfall-triggered landslides is evaluated in many cases by empirical threshold methods or multivariate statistical techniques (Caine, 1980; Corominas, 2000; Fan *et al*,. 2003). However such approaches may fail because of the variety of landslide types in an area responding in a different way to the meteorological input. Moreover historical datasets connecting meteorological data to failure incidents to derive such thresholds are rare (Coe *et al.*, 2004). It is obvious that for the assessment of meteorological thresholds, shallow landslides (1-2 m) require different meteorological information than deeper landslides. For deeper landslides a large windows of antecedent daily precipitation over weeks or months, including daily evapotransporation will determine the threshold for failure while for shallow landslides one has to consider only a few rain events or even one, with known intensity and duration, to forecast failure (van Asch *et al.*, 1999). It is therefore obvious that the assessment of a simple meteorological threshold based on empirical calibration of one or two meteorological parameters is not possible for an area with a variety of landslide types (Malet *et al.*, submitted).

Consequently, the hydrological system has to be studied carefully for relevant hazard analysis and the ground-water flow field has to be modelled by superposition of a steady component and a transient component (van Asch *et al.*, 1999; Iverson, 2000).

### Challenges and future research directions

Hydrological model results are very sensitive to the steady-seepage initial conditions which are probably the most significant input for the modelling effort. Obtaining accurate initial conditions requires a significant number of field observations possibly supplemented by steady-flow modelling to extrapolate between observations (Iverson, 2000; Wilkinson *et al.*, 2002) but also to critically take into account the role of the unsaturated zone with its highly non-linear behaviour and its buffering capacity (Ng & Shi, 1998; Bogaard & van Asch, 2002).

For a good forecast of the pore pressure distribution and time delay of the meteorological input signal, it is necessary to use coupled unsaturated-saturated infiltration models assuming that the flow is in the linear range for Darcy's law and the hydraulic diffusivity is approximately constant. Infiltration in unsaturated soils is more complex than in saturated soils since the initial degree of saturation of the soil profile and the initial negative pore pressures (e.g. matrix suction) control the hydrologic conductivity and the quantity of water required to reach full saturation (Torres *et al.*, 1998). Relationships between degree of saturation, matric suction and hydrologic conductivity are not simple either (Fredlund & Rahardjo, 1993). Wang & Thomas (2000), Bogaard (2002) and Bogaard & van Asch (2002) have shown that blab la bla…. Theo: to complete, introduce Fig. 7.9. of Thom's thesis and explain it.

Fig. 5 – Fig. 7.9. Thom's thesis

Development of pore pressure in initially unsaturated soils is also complicated by the influence of the local topography (e.g. stress relief; Bromhead & Dixon, 1984; Hulla et al,. 1984; Torres *et al.*, 1998), the influence of the vegetation on water losses by evapotranspiration (Eigenbrod & Kaluza, 1999) and the influence of preferential flows (fissures, desiccation cracks, root holes, animal burrows). Especially, the complex geomorphogical structure of landslides and the presence of fissure systems may result into complex and inter-connected hydrological subsystems. For instance the stability of deeper landslides in varved clays in France is controlled by a perched groundwater table in the morainic top layer feeding the deeper fissures (van Asch *et al.*, 1996). From a detailed analysis of the

Super-Sauze mudslide, Malet *et al.* (2005) have shown that incorporating a conceptual (grey-box) model of fissure flow in a physically-based infiltration model describing matrix fluxes can lead to more accurate simulations of the soil moisture contents in the unsaturated zone (Fig. 6).

Fig. 6 – Influence of fissure flow in simulating rainfall infiltration in the unsaturated zone of the Super-Sauze mudslide for two periods (modified from Malet at al., 2005). The graphs indicate the observed and simulated variations of volumetric soil moisture contents at two depths (-0.56m; -1.08m) with or without introducing fissure flow. The matrix infiltration is simulated with a physically-based model; the fissure flow is simulated with a conceptual model of direct bypass of the rain water to the bottom of the fissures.

But it is still extremely difficult to quantify the influence of preferential flows on soil stability, especially because the architecture of the fissures and the flow processes in the fissures are difficult to detect. The challenge is to describe, experimentally and numerically, the hydraulic behaviour of water in fissures and the interaction with the soil matrix, and to upscale these concepts to the slope scale. Interesting information can be gained from the monitoring of electric signals (e.g. the soil electrical streaming potential) by innovative geophysical techniques at the laboratory scale as well as in the field (Sailhac *et al.*, 2004), and from the use of chemical tracering techniques to quantify the water fluxes at different scales (Di Pietro *et al.*, 2003; Weiler *et al.*, 2003; Bogner *et al.*, 2005).

The hydrology of low permeability deposits (especially clays) has also to be studied. The water flows in saturated clayey soils are controlled by the swelling/consolidation coefficients of the material, and possible presence of layering like for instance in stiff varved clays (Nieuwenhuis, 1991). Accordingly, the seasonal variations of pore pressures at the boundaries of a clay deposits may not be entirely reflected in the entire deposit (Picarelli *et al.*, 2000).

As a result, the use of our modelling capabilities is limited to modelling hypothetical scenarios. A better understanding of the hydrological processes and of an accurate control of the initial conditions is needed to forecast occurrences of failure induced by climate and land use changes (Bonnard & Noverraz, 2001; Bogaard & van Asch, 2002; van Beek, 2002). It is also apparent that a lengthy period of groundwater observation is likely to be necessary to experience a range of rainfall and piezometric conditions including high rainfalls after long droughts for example.

## 5. Post-failure behaviour: slow and rapid development of the processes

### *Slow developments of slope movements and the potentials for catastrophic reactivation*
Theo: to review and structure:
*Slow moving gravitational processes*
(Rens comment: *This should be formulated more pressingly, in general terms rather than start with "We...*)
We have to ask our selves whether the current models have the capacity to forecast the transient effects of slow moving landslides. In fact we are mainly interested in the forecasting of periods of crises of these slow moving landslides. The parameterisation of hydrological and geomechanical factors obtained from field and laboratory tests , like for example viscosity are not sufficient to describe the moving pattern of these slow moving landslides. Additional mechanisms play a role in the field, which are difficult to simulate.(van Asch *et al.* in press). We are confronted with complex hydrological systems consisting of fissure and matrix flow regimes as described above. These fissure systems may change during movement due to a change of the geometry. This may drastically change the rate in discharge (and drainage) of the groundwater body (Corominas *et al.*, 1999; Malet *et al.*, 2005) + Hydro-mechanical modelling La Frasse (Bonnard, Tacher,, etc) + Picarelli

Slow moving landslides are not rigid moving bodies. Zones of compression and extension will be generated caused by heterogeneity of the moving pattern. This will create undrained loading effects leading to the generation of excess pore pressure (Giusti *et al.* 1996, Picarelli *et al.* 1995). The hysteresis in the velocity pattern during a rising and falling limb of the groundwater which is observed on slow moving landslides (Leroueil *et al.*, 1996; Malet, 2003 ) can be explained by this mechanism of undrained loading (van Asch, 2005).

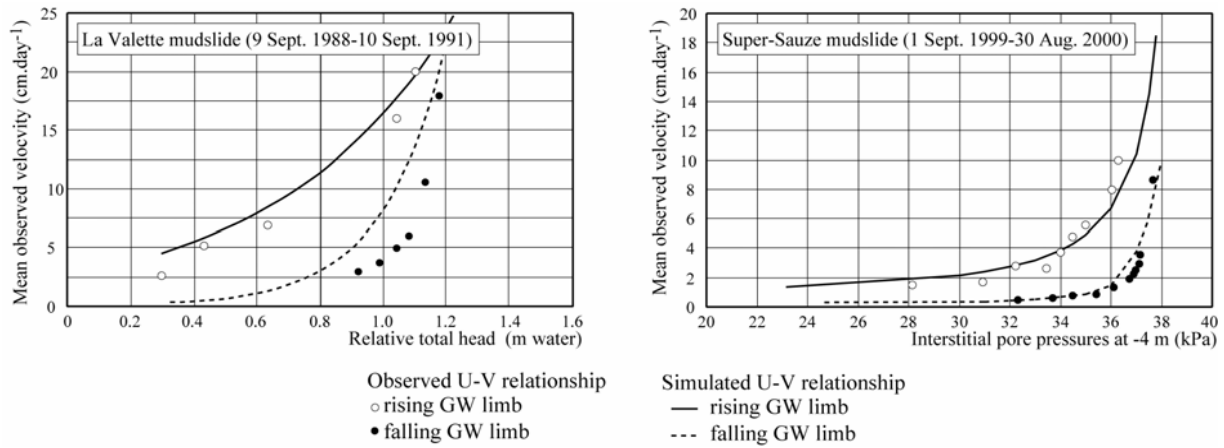

Observed U-V relationship
○ rising GW limb
● falling GW limb

Simulated U-V relationship
— rising GW limb
--- falling GW limb

Fig. 7 – Observed and simulated hysteresis in the velocity pattern during a rising and falling limb of the groundwater in two slow-moving mudslides due to compression and undrained loading (La Valette, South French Alps, Super-Sauze, South French Alps [modified from Malet (2003) and van Asch (2005)].

There is also a possibility of a spatio-temporal trend in the change of the strength due to the rheologic properties of the material. There may be a loss of shear strength during movement and extra strength (higher viscosities) in the field may be gained through the development of negative excess pore pressure. (Keefer and Johnson 1983) contribute strength development during movement not to the pure viscosity but hundred percent to pore pressure effects developed by a porous elastic solid, sliding over a wavy rigid surface. Compression may develop at the proximal sides of bumps in slip surfaces and extension generating negative pore pressures at the distal sides. van Genuchten and van Asch (1988) found similar feedback mechanisms in intermittently sliding blocks of the La Mure landslide. The movements of the blocks showed a stepwise character pointing to a "slip-stick" mechanism. Movements were generated by a rise in pore pressure due to infiltrating precipitation. A stop of the movement did not coincide with a lowering of the groundwater. It was ascribed to irregularities in the slip surface (large stones, boulders) forming cavities during movement behind these boulders, creating large but very local suctions reducing the mean pore pressure in the slip surface. In landslides with a intermittent moving pattern, strength regain by consolidation may occur also during period at rest (Nieuwenhuis, 1991; Bertolini and Pellegrini, 2001; Angeli *et al.*, 2004) (Could be condensed)

Fig. 8 – Fully coupled hydro-mechanical simulation of an acceleration of the Super-Sauze mudslide, South French Alps (GefDyn finite-element code) (Malet, 2003). The shear zone is bounded by a slip surface which has been simulated by an interface element. The soil geomechanical behaviour is assumed to be an elasto-visco-plastic Hujeux material while the bedrock below the sliding surface is a stiff stable material. The overall safety factor of the slope at the beginning of the simulation is approx. 1.2, very close to failure. A pore pressure increase lasting 5 days (from 56 to 62 kPa) and corresponding to observed data is simulated. As the pore pressures gently raise, the state of stress in the mudslide changes and a part of the soil mass tends to accelerate (stage A). The consequent new situation is then simulated by an undrained (short-term) analysis allowing the associated pore pressure to equalize while the deformation is still continuing at a constant rate. Combination of continuing infiltration and soil deformation lead to a second acceleration (stage B). This local failure is rapid enough to cause catastrophic excess pore pressure with some delay after the time of the second acceleration, causing the global failure of the secondary scarp of the mudslide (stage C). Such type of analyses allow to analyse the interaction between landslide movements and the development of pore pressures induced by both infiltration and undrained compression of the soil.

*Transient behaviour and potential for fluidisation after failure.*
One of the major problems is to forecast the probability that sliding mass movements after failure fluidise and mobilize into dangerous rapid flows, which has much larger impact areas. Different mechanisms has been identified which explain this dangerous transient behaviour of landslides. These phenomena have been observed many times in loosely packed material, which contracts during shear failure inducing a catastrophic rise in pore pressure causing fluidisation (Yoshimi *et al.*, 1989; Anderson and Reimer, 1995; Iverson *et al.*, 1997; Fuchu Dai *et al.*, 1999). However, liquefaction phenomena have also been observed in more compacted soils, which dilate during shear failure. The physical relevance for more compacted soils, especially during initial movement, has to be further investigated in the laboratory using small-scale experiments (The processes that govern liquefaction of compact soils are still poorly understood and should be the subject of fundamental research). Move this up and close with the previous sentence: It can be generated by simple undrained loading caused by a changing stress field during initial failure (Baum and Fleming, 1991; Picarelli *et al.*, 1995; Giusti *et al.* 1996; Klubertanz *et al.*, 2000; Comegna & Picarelli, 2005; van Asch *et al.*, in press). Another possibility is that a geometrical change at the toe of the landslide may increase the effective stress so that the material may pass the critical state line and transfer from a delative in a contrative state (Reimer, 1992, Gabet & Mudd, in press). Initial porosity is crucial for the development of rapid flows through liquefaction in sliding material (Iverson *et al.*, 2000) and field investigations on liquefaction combined with geomechanical analyses of the involved material has to reveal more insight under what conditions more compacted material with lower porosity will liquefy.

Fig. 9 – Geomechanical simulation of the failure of the secondary scarp of the Super-Sauze mudslide, South French Alps, by undrained loading assuming an elasto-plastic material (Flac 2D finite-difference code). The effect of undrained loading can be modelled in relatively simple terms using numerical models that include fluid-mechanical interactions like FLAC. Starting from a critical slope with a safety factor F=1.0, incipient deformation leads to undrained loading that affects the short-term stability negatively. Movements translate into a noisy but gradual increase in pore pressure that in turn leads to an ongoing deformation than that required to accommodate the unbalanced force within the grid. Once a new equilibrium is maintained, pore pressures stabilise at a higher level due to the cumulated displacement and may dissipate ultimately to return a new long-term stability for the slope.

### JPM Rapid development (fast gravitational flows)
5.2.1 State-of-the-art in understanding the mechanisms
Rapid gravitational processes, like mudflows and debris flows, are very frequent and they are the most dangerous type of landslides. In fact there are a couple of interrelated processes, which has to be considered carefully in our modelling attempts. Important issues, which has to be dealt with, are the assessment of meteorological triggering thresholds and the hydrological triggering mechanisms at initial failure in the source area, the volumes which are mobilized into a flow, the amount of erosion and transport of the flow material and sedimentation and the assessment of run-out distance by preference in a 2D pattern. The modelling of these quantities meets a lot of difficulties.

### 5.2.2 Examples of models, challenges & future research directions
Flow run-out forecasting can be done by black box modelling on the basis of former incidents to construct maximum friction lines, which determine run-out distances or friction lines with variable angles related to environmental factors to create a GIS zonation of impact probability. However these types of empirical analyses require a lot of data, which are not available in most cases. (van Westen *et al.*, 2005). Therefore much attempt has been given to the development of physical based models.

For hazard and risk analyses of rapid gravitational flows it is important to estimate temporal frequencies and thus meteorological thresholds for triggering. The estimate of run out distances which is in the first place controlled by the amount of mobilised volume of material is of equal importance. In case the material is delivered by sliding material the amount of volume can be estimated by classical stability analyses where an estimate can be made of critical slip surfaces and hence the volume, which failed. However that requires also an estimate of groundwater flow and heights which is delivered by rain or (and) run off water.

The next step is to estimate the volume of material which will liquefy and transform into a debris flow. This depends on the mechanism of liquefaction, which may be caused by compaction of material during shearing, undrained loading or (and) geometrical deformation during initial failure (see above). There are other external mechanisms and processes involved, which make it difficult to assess volumes and meteorological thresholds. Apart from failure of in situ soil material, debris material accumulated in gullies in the source area, can be remobilised and turn into debris flows. Triggering of this material can take place by run off water infiltrating the debris mass or by entrainment of this debris by heavily loaded run-off water. It is difficult to forecast which process is dominant. (Blijenberg, 1998). Also flow material can be delivered in the source area by erosion in steep gullies which will heavily load the runoff water with sediment (Hessel, 2002), collapse of gully walls during high discharges of water or hyper-concentrated flows and scouring of in situ material by debris flows in the run out track (Chen & Yan, 2003). We do not know until now what precisely are the processes which deliver high sediment concentrations on very steep slopes with high amounts of runoff water. Is it only detachment of grains by run off water which has a high transport capacity? Or is it a combination of run off erosion and micro slumping/flowing , which causes a high amount of sediment detachment resulting in a hyper concentrated flow (Blijenberg, 1998)? The stochastic character of these processes make it difficult to forecast the right type of processes involved. More detailed field observations and laboratory experiments are necessary to improve our modelling capacity of these processes (Remaitre *et al.*, debris flow)

For an estimate of impact on infrastructures and buildings, it is important, apart of volumes, to estimate velocities. These velocities are largely dependent upon the material intrinsic properties, which are highly variable, and likely to change during the flow itself (Savage & Hutter, 1991; Takahashi, 1987; Rickenmann, 2000; GDR Midi, 2004; Denlinger & Iverson, 2004; Iverson *et al.*, 2004). Advanced numerical models are available to model velocities and run-out distances (Chen, 1987; Hungr & Evans, 1996; Laigle &Coussot, 1997; Denlingr & Iverson, 2004).These models require a lot of parameterisation to back calculate several rapid gravitational movements with a good accuracy. Developing better models, improving the quality and number of variables to be fed into these models are therefore necessary tasks. Better rules for calibration and better methods for the measurements of important parameters are also necessary (Iverson, 2003).

Therefore available data on several mountain torrents, experiencing a high activity of events, has to be explored to correlate different kind of material properties (including the volumetric distribution of flowed material along the track) to the rheological properties of flow in order to select the appropriate flow simulation models for a better estimate of velocity and run-out distances. (Bardou *et al.*, 2003; Coe *et al.*, 2003; Malet *et al.*, 2004; van Asch *et al.*, 2004)(figure Rio Figure PC raster)

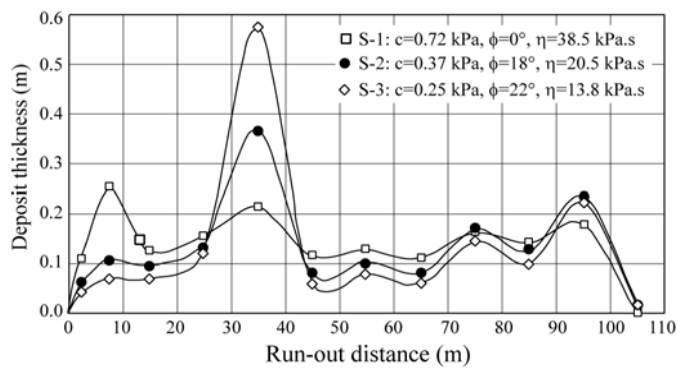

FIG. 10 – 1-D runout modelling of 15 m³ volume of debris material along a run-out track at the Super-Sauze mudslide, South French Alps) showing the effect of material properties, for a Bingham material (Scenario S-1) and Coulomb-viscous material (Scenario S-2 & S-3) (modified from van Asch *et al.*, 2004).

However a better selection of appropriate material parameters to feed these models for calibration and validation, can be made if we are able to measure velocity, depth, discharge pore pressure distribution, grain size distribution and concentration during the flow event (Lavigne, 2004) .Different measuring techniques have to be developed, extended and tested to measure under difficult circumstances important flow characteristics during the event. Different kind of sediment trap techniques to measure sediment sediment concentration during the event are of utmost importance  This has to be combined with systematic rheological tests in the laboratory with material of different sediment and rock concentration. (Ancey, 2003). The rheological characteristics obtained in the laboratory must be compared with back analyses of flows with comparable material on sites where also velocity depth and concentration of material could be measured. Geophysical techniques to measure the hydro dynamic characteristics of the flow must be extended (Lavigne & Suwa, 2004). Video recording proved to be very useful to study the dynamic behaviour by measuring flow velocities, concentration and kinematic behaviour of the blocks. (Suwa, 1988; Zhang and Chen, 2003).

The amount of measuring sites where frequently debris flows occurs must be extended for different geomorphological and material conditions as well as the amount of experiments in smaller and larger flumes (Iverson, 2003). However given the extreme difficult parameterisation of sophisticated models for debris flow, we have to continue with developing and testing the forecast of simple robust models, which require a limited amount of parameters (Hungr, 1995; Rickenmann, 2000; De Joode & van Steijn, 2003).

FIG. 11 – 2-D runout modelling of 5000 m3 volume of debris material on an alluvial fan showing the effect of material rheology (viscoplastic Bingham rheology, frictional Voelmy rheology) on the spatial distribution of material. The black lines represent the elevation curves of the DEM and the coloured lines represent the thickness of the material deposits.

Hazard and risk assessment for debris flows is of great importance on alluvial fans (figure PC raster) where a great part of the vulnerable habitation and infrastructure can be expected. In order to model properly and to forecast the 2D run-out pattern of the flow material on these fans, we need detailed topographic information from these depositional areas. This is a problem because of the lack of accuracy of the available DTM's and the stochastic changes in topography during the depositional process. Nowadays an improvement of the precision of our DEM's can be expected with laser scanning techniques such as LiDAR. This airborne method, forms a new and important tool for detailed topographic mapping which will be beneficial for many aspects of landslide hazard modelling (Norheim *et al.*, 2002)

## 6. Extent of application: analysis of landslide hazard at the regional scale

When studying potential hazards outside the boundaries of existing active and monitored slope movements, simulations based on probabilistic models as well as event-tree methods (Oboni 1984; Cherubini & Masi, 2002; Dai & Lee, 2003; Hsi & Fell, 2005) are the tools of preference to extent the knowledge gained at the scale of individual slopes to a large number of slopes, as well as to estimate probabilities and magnitudes of slope movements (Aleotti & Chowdhry, 1999; Haneberg, 2000; Wong, 2005).

  Many statistical techniques have been developed and applied successfully to landslide susceptibility assessment and mapping in the last ten years using bivariate or multivariate approaches, probabilistic approaches (like Bayesian inference or logistic regression) and artificial neural networks approaches (Carrara, 1983; Carrara *et al.*, 1991; Fabbri & Chung, 1996; Guzetti *et al.*, 1999; Ermini *et al.*, 2005).

Applications on field data have shown that in some cases quite good spatial predictions can be made using those models and relatively small number of conditioning variables (Coe *et al.*, 2004; Zêzere *et al.*, 2004a; van den Eeckhaut *et al.*, 2006). Nevertheless these techniques lack the support and skill to evaluate temporal probabilities, transient effects and long-term changes on landslide activity.

Various types of probabilistic assessments based on historical and Quaternary records are used by paleo-seismologists and paleo-hydrologists in determining earthquake and flood hazards, but probabilistic assessments of future slope movements occurrence are rare, and the published assessments provide only a ranking of terrain units in terms of susceptibility (e.g. a time-independent, spatial, distribution of future events; Chung & Fabbri, 1999). Time-dependent assessments are difficult to apply because most available event records cover short periods and small geographic areas. Also, records often do not contain information on the date of occurrence or reactivation of the slope movement, the volume mobilized or event the type of movement (Ibsen & Brunsden, 1996; Hungr *et al.*, 1999). Often, different landslide types are merged into one training dataset which obscures the influence of different controlling factors even further (Malamud *et al.*, 2004; van Westen *et al.*, 2005). Furthermore temporal information on the meteorological or seismic triggering events is scarce in most cases. These variables are needed to set up reliable magnitude-frequency curves at the regional scale (Guthrie & Evans, 2005).

Therefore a challenge in the forthcoming years is to use new strategies to add to the already available susceptibility assessments a temporal dimension in order to produce real hazard maps. The use of deterministic (physically-based) methods (Dietrich *et al.*, 2001; Chen & Lee, 2003; Savage *et al.*, 2003) in combination with probabilistic statistical techniques should theoretically be able to overcome these problems (van Beek, 2002; Casadei *et al.*, 2003) assuming that detailed spatial and historical databases are available. Several research directions are already suggested.

Coe *et al.* (2004) analysed a very detailed database of rainfall-triggered landslides in Seattle, (Washington, United States) using a Poisson statistical model to estimate the probability of future occurrence of individual landslides, and a binomial statistical model to estimate the probability of having a group of one or more landslides within an individual year. Each model application produces a map showing landslide densities (number of landslide per given area) or landslide cluster densities (number of years with one or more landslides) as well as mean recurrence intervals and exceedance probabilities.

Dussauge-Pessier *et al.* (2002) and Hantz *et al.* (2003) used a multi-scale approach to derive frequencies of rock falls based on volume ranges in the Chartreuse Massif (French Alps). These frequencies allow transforming the spatial probabilities of the potential location of the unstable masses into failure (temporal) probability, and thus hazard. Spatial probabilities are calculated through statistical and geo-mechanical analyses, and temporal probabilities are calculated through inventories at different time scales. Different scenarios concerning the volume can then be considered with their respective probabilities. Similar approaches have been used by Guzzetti *et al.* (2003, 2004).

Zêzere *et al.* (2004) integrated the spatial and temporal probability of shallow landslide occurrences in the Fanhões-Trancão area in the North of Lisbon (Portugal). The authors used logistic regression algorithms (over unique conditions terrain units) on a landslide inventory which was classified by type and time period to obtain spatial probability estimates. They combined these spatial probabilities with the known return periods of rainfall-event that triggered the different landslide types. This combination results in an integrated spatio-temporal landslide probability map.

When information is scarce about temporal distribution, thresholds for failure/reactivation of a certain landslide type in a given area, and magnitude, scenario modelling can provide significant information and trend. van Beek & van Asch (2004) used ….. Theo: bla bla bla introduce and explain Fig. 12 in 2-3 lines

FIG. 12 – Landslide occurrence observed over the period 1973-1994 in the Alcoy catchment, Southeast Spain (12a) and simulated maximum probability of failure (12b) (modified from van Beek and van Asch, 2004).

Malet *et al.* (submitted) proposed to use Probability Density Functions (PDFs) of rainfall and groundwater heights to investigate stochastically the failure occurrences within a slope with a deterministic coupled hydrology-hillslope stability model. The model runs are performed for many slope geometries, many soil characteristics and many initial conditions. This approach delivers information about the magnitude (e.g. volumes of material able to fail) and the thresholds for failure that can be crossed with a probabilistic susceptibility map. The approach necessitates detailed data on soil thicknesses which may be difficult to obtain (Terlien *et al.*, 1995; van Beek & van Asch, 2004). PDFs can also be used to handle the variability of the material characteristics (Haneberg, 2000; Hamm *et al.*, 2006).

The same type of approach is interesting to investigate runout frequencies and magnitudes of landslides in the absence of documentation of former events (volume involved, landslide travel distances). Malet & Begueria (submitted) proposed a methodology to compute the characteristics of low-frequency debris flows through Monte Carlo techniques combining a deterministic 2D flow model and a probabilistic description of the model input parameters.

Magnitude-frequency curves of torrent discharges and a multivariate distribution function of geomechanical parameters (density, yield stress, viscosity) from various well-documented torrents are used to generate a random distribution of input parameter vectors. Many model runs are then performed using the randomly generated input values, and the spatial probability of occurrence (e.g. probability of a pixel to be affected by material deposition) is calculated. The degree of hazard, expressed as a time probability or a recurrence interval, is then computed by combining the magnitude/frequency of the discharge and the probability of occurrence. A schematic representation of the methodology is described on figure 13.

FIG. 13 – Probabilistic assessment of debris flow hazard on an alluvial fan, by combining Probability Density Function of input model parameters, Monte-Carlo simulations and model runs. Examples of Monte-Carlo simulations of debris flow heights (Malet & Begueria, submitted).

It is also sometimes practical to simulate the sequence of events (using an event-tree and expert opinion) which may lead to an individual slope failure, and thus estimate a frequency of failure. This approach, very often used for earthquake hazard assessment, stands on the observation of real cases and on some conclusions of how a slope movement would be initiated and would behave. Hsi & Fell (2005) used this approach to assess the hazard associated to a coal cliff in Australia. This approach is promising since at any node of the tree, conditional probabilities could be assigned to those events coming from the former node, and the probabilities can then be summed up.

Comment Rens: There is a good book by Mike Leigh on landslide hazard risk and probabilities. OK BUT WHAT? Do you have the title of this book; I did not find it on WWW

## 7. Role of geomorphology in improving our modelling performance

To quantitatively assess slope movement hazards, simulations based on both long– and short-term modelling of slope evolution, combining many sources of knowledge, and conducted with probabilistic approaches are necessary tools to handle the variability of the controlling factors as the uncertainties of their measurements. Baynes & Lee (1998) discuss the role of geomorphology in landslide hazard assessment.

On long time scales, geomorphological analyses and modelling of slope evolution deliver quantities like the weathering rate of the materials (from hard rocks to weathered less resistant soils), the denudation/deposition rate of the soils (linked to the progress in depth of the weathering front and to the transport of sediment on the slope) and the rate of uplift/incision of the landforms by rivers or glaciers. Long-term hillslope modelling and reconstruction of landscape evolution can help to quantify the temporal evolution of predisposing factors and among them slope angle, soil depth and soil shear strength (Ahnert, 1987; Montgomery & Dietrich, 1994; Hovius *et al.*, 1997; Kirkby, 1998, 2003; Perini *et al.*, 2001), and identify the state of activity of landslides (Cenderelli & Kite, 1998; Caine & Swanson, 1999; Korup et al., 2004; Claessens et al., 2006). Nevertheless, most of the slope development models are still not detailed enough to forecast the evolution of these factors towards instability on large spatial and temporal scales (Trustum et al., 1999). A major obstacle when assessing rates of landslides is the difficulty of obtaining data that are relevant over medium to long time scales (Crozier, 1996; Martin et al., 2002). Consequently, detailed chronological analyses of landslide sediments (preserved in swamps or in lakes) by dating techniques, combined to erosion/sedimentation modelling to calculate landslide sediment volumes and/or sediment yields from a catchment are necessary tasks. The physically-based and spatially distributed models proposed by Burton & Bathurst (1998) and by Claessens et al. (2006) are interesting tools able to estimate runout distance from hillslope geometry and to provide maps of soil redistribution.

On shorter time scales, geomorphological observations can help to understand the type and the mechanics of movement, but this stage in the investigation is often ignored. Geomorphology may therefore reveal the complexity of real-life landslides and thus the inevitable shortcomings of abstract models. Specific geomorphologic signs enable us to reconstruct the type of processes involved, may reveal the sequence of kinematics during failure (Geertsema et al., 2005a), which is important for the selection of relevant hypotheses in the modelling of the system (Dikau et al.,1996). Distinctive geomorphological features for the identification of landslides can be found in the source area, in the development area and in the accumulation area. In the source area, the geometry of the crown (upper limit of the landslide complex) and the slope of the main scarp (steep slope created by the displaced material), and the type of deformation of the topographical surface (back tilting slopes forming ponded lakes) are relevant indicators to identify the geometry of the failure. The topography of the development area (e.g. the main body), its degree of disturbance, the pattern of ridges and cracks and the contour of the main body (elongated or strong lateral spreading) are relevant indicators to understand whether the material is sliding or flowing. Finally, in the lower part, the form and steepness of the toe (lower limit of the landslide), the pattern of cracks and ridges are also indicators. The freshness of cracks, striation lines and disrupted topography and the stage in the vegetation growth are field evidences of the activity of the landslide complex (Crozier, 1986). Two interesting examples of cascading sequence of failure, expressed by geomorphological indicators, are provided by the La Valette landslide (South French Alps, Fig. 14a) and the Muskwa landslide (North-West British Columbia, Canada, Fig. 14b).

Fɪɢ. 14 – Geomorphology of complex landslides. (15a) Complex La Valette slump-mudslide in the French South Alps; (15b) Complex Muskwa slide-earthflow in North-West British Columbia, Canada. The La Valette landslide complex shows a steep scarp with a backward tilted block (initial slump) in the upper part, a less disturbed surface showing subsidence and movement parallel to the slope in the middle part, and an elongated mud track with a clear lobate form and argued transfer ridges in the lower part. The Muskwa landslide bla bla bla Description Theo.

Furthermore, geomorphological observations may support the conceptualization and evaluation of the process-based models (van Beurden, 1997; Remaître *et al.*, 2005; Geertsema et al,. 2005b). A thorough investigation and monitoring programme of an individual landslide needs therefore to combine geomorphological, geotechnical, geophysical and hydrological analyses (Bogaard *et al.*, 2001) as outlined by figure 15.

Fɪɢ. 15 – Multi-source strategy of investigation and monitoring of an active slope movement.

The state-of-the-art papers provided in this Special Issue address relevant recent technological developments made on the identification of landslide displacement by remote-sensing techniques

(Delacourt et al., 2006), on the identification of landslide geometry and internal structure by geophysical techniques (Jongmans & Garambois, 2006) and on the analysis of the hydrological system of landslides by hydrogeochemistry techniques (Bogaard et al., 2006). These issues are therefore not discussed in this paper.

   Attention is nevertheless put on the needs for relevant and accurate topographical information both for understanding the processes and for testing the performance of the models. These techniques are useful to support geomorphological investigations in order to identify the micro-relief of dormant or stabilised landslides which are always places of new activity and thus an important weighing factor in regional hazard assessments. They are also critical to identify the boundaries of the phenomena and to elaborate high resolution DEMs. Useful techniques for landslide hazard assessment are stereo-photogrammetric analyses of aerial air photographs (Herrmann & Weber, 2000; Chandler, 1999; Casson et al., 2003), optical or radar remote-sensing (Massonnet & Feigl 1998; Kimura & Yamaguchi, 2000; Squarzoni et al., 2003; Delacourt et al., 2004) or ground-based techniques like dGPS (Malet et al., 2002) and terrestrial SAR interferometry (Tarchi et al., 2003).

**Conclusions**
Theo: edit again the conclusion section …. I did not check it for the moment.
It is rather difficult nowadays to make further progress in our modelling of landslide hazard and risk at different scales in the pre-failure and the post-failure stage. There will be gains but small!
The most difficult issue is to model over a long-term period the preparatory path to failure. In rock material we are able to model and understand in detail the development of fissures in a physical way, but it is difficult to detect and follow the evolution of the architecture of fissures in depth at the hill slope scale. This hampers the calibration and validation of models which describe the stress strain history before failure. Rate of weathering and the evolution to critical soil depth and slope profiles are the main preparatory agents for slope failure in soil material. There are models which describe these slope development processes, but they are seldom used to explain and forecast the temporal activity of landslides. Modelling of slope deformation on the short term just before failure is difficult because of external disturbances with a stochastic character.
The most important trigger for failure on the short-term time scale and for the reactivation of landslides is the hydrological system. In most cases the hydrological system in landslides is poorly described. In the first place less attention is given to the unsaturated zone, which controls the rate of groundwater supply and which it self is controlled by the evaporation capacity of the vegetation mantle. Ignoring the role of the unsaturated zone makes it difficult  to forecast effects of land use and climate change on slope stability. The hydrological system in landslides may be rather complex due to the presence of fissure systems. The difficulty is to detect the (changing) structure of these fissures and their connectivity and to quantify the water fluxes in these fissures and the exchange of water with the matrix system.
Different mechanisms can play a role when sliding material liquefy and transform into a flow. The amount of material, which liquefy is a major controlling factor for the run-out distance of these flows. For compacted material we are not quite able to forecast whether fluidisation might occur and how much material will pass into a flow. Also we do not understand quite well the mechanisms of the processes on very steep slopes with high amounts of runoff water, contributing to sediment delivery of hyper concentrated flows.
A major problem in the run-out modelling of the debris flows is the selection of the right rheologic properties of the material, which often may change during the run-out. We have to extent our search for morphological indicators, our monitoring activity, our techniques for taking samples during the run-out, and our experiments in the laboratory flumes in order to understand the rheological behaviour under different geomorphological circumstances and for different materials. An increase in our knowledge and experience of debris flow propagation must enable us to make a good balance between the increased details of the process description vs. the parameterisation load.
For slow moving landslide bodies the major challenge is to forecast periods of crises. This requires a detailed understanding of the factors controlling the moving pattern. Rheological properties obtained from laboratory test are not enough to model the velocity. A number of factors operating at the field scale have to be included. They are related to the generation of compressive and dilative stresses during movement, causing excess pore pressure. Also there are changes in strength characteristics due to consolidation in period at rest and reactivation, and there may be increase in strength due to pore pressure effects caused by the irregularities in the slip surface. The transient character is also influenced by changes in the geometry of the landslides and especially possible changes in the geometry of the fissure system, which may completely change the hydrological water balance

Landslide hazard and risk assessment at the catchment scale needed for planning purposes and cost-benefit analyses, requires information on the temporal impact frequency of these processes. This information can seldom be delivered due to the lack of historical data. It is a challenge for the future to get this information from physically based hydro-mechanical models. There is a lot of work to do, to integrate these models in our hazard zonation maps showing the spatial probability of landslide processes. Investigations have to be carried out on how far the physically based models are representative for the range and type of landslide processes in a certain area and on the required level of parameterisation for reliable modelling of landslide frequency.